\pdfoutput=1
\documentclass[conference,a4paper]{IEEEtran}
\IEEEoverridecommandlockouts

\usepackage{graphicx}
\usepackage[caption=false,font=footnotesize]{subfig}
\graphicspath{{./figs/}}
\usepackage{microtype}
\usepackage{textcomp}
\usepackage[table]{xcolor}
\usepackage{tikz}
\usetikzlibrary{arrows.meta}
\usepackage{booktabs}
\newcommand{\experimenttablefont}{%
  \fontsize{6}{6.6}\selectfont
  \renewcommand{\arraystretch}{1}%
  \setlength{\tabcolsep}{0.4pt}%
  \setlength{\aboverulesep}{0.8pt}%
  \setlength{\belowrulesep}{0.8pt}%
  \setlength{\cmidrulesep}{0.8pt}%
  \setlength{\heavyrulewidth}{0.5pt}%
  \setlength{\lightrulewidth}{0.3pt}%
  \setlength{\cmidrulewidth}{0.3pt}%
}
\usepackage{multirow}
\newsavebox{\timinghistogrambox}
\newsavebox{\timingablationbox}
\newlength{\timingfigureheight}
\usepackage{threeparttable}
\usepackage{etoolbox}
\makeatletter
\patchcmd{\@makecaption}{\scshape}{\upshape}{}{
  \PackageError{ReSynP3D}{Table caption font patch failed}{}
}
\patchcmd{\@makecaption}{\\}{\nobreakspace}{}{
  \PackageError{ReSynP3D}{Table caption separator patch failed}{}
}
\patchcmd{\@makecaption}{\addvspace{0.5\baselineskip}}{}{}{
  \PackageError{ReSynP3D}{Table caption spacing patch failed}{}
}
\def\@IEEEtablecaptionsepspace{\vskip 2pt\relax}
\newcommand{\unindentedfootnote}[1]{%
  \begingroup
  \long\def\@makefntext##1{\noindent\hbox{\@makefnmark}##1}%
  \footnote{#1}%
  \endgroup
}
\makeatother
\usepackage[colorlinks=true,citecolor=blue,linkcolor=blue,urlcolor=blue]{hyperref}
\usepackage{xurl}
\usepackage{balance}
\usepackage{amsmath,amsfonts,amssymb}
\usepackage[linesnumbered,ruled,vlined]{algorithm2e}
\usepackage{cleveref}
\crefname{algocf}{Algorithm}{Algorithms}
\Crefname{algocf}{Algorithm}{Algorithms}
\crefname{algorithm}{Alg.}{Algs.}
\Crefname{algorithm}{Algorithm}{Algorithms}
\Crefformat{figure}{Fig.~#2#1#3}
\crefformat{figure}{Fig.~#2#1#3}
\Crefname{figure}{Fig.}{Figs.}
\crefname{figure}{Fig.}{Figs.}
\Crefname{table}{TABLE}{Tables}
\crefname{table}{TABLE}{Tables}
\Crefname{section}{Sec.}{Secs.}
\crefname{section}{Sec.}{Secs.}
\Crefname{equation}{Eq.}{Eqs.}
\crefname{equation}{Eq.}{Eqs.}
\usepackage{enumitem}

\newcommand{\ours}{ReSynP3D}

\begin{document}

\title{\fontsize{23}{27}\selectfont ReSynP3D: Bridging Logic Resynthesis and Tier Repartitioning during 3D~IC Placement Optimization}

\author{
    Shuo Ren\textsuperscript{1}, \quad
    Shixi Chen\textsuperscript{2}, \quad
    Yaohui Han\textsuperscript{1}, \quad
    Fangzhou Liu\textsuperscript{1}, \quad
    Mingjun Li\textsuperscript{1}, \quad
    Libo Shen\textsuperscript{1}, \\
    Xuanqi Chen\textsuperscript{2}, \quad
    Rongliang Fu\textsuperscript{1,*}, \quad
    Bei Yu\textsuperscript{1}, \quad
    Tsung-Yi Ho\textsuperscript{1}\\[2pt]
    \textsuperscript{1}The Chinese University of Hong Kong \qquad
    \textsuperscript{2}The Hong Kong University of Science and Technology
}

\maketitle
\pagestyle{empty}
\thispagestyle{empty}

\begin{abstract}
As Moore's law slows, 3D ICs offer a promising path to continued improvements in performance and energy efficiency.
In 3D ICs, Boolean structure and tier assignment jointly determine
critical-path delay, yet fixing the logic structure before placement leaves
cross-tier restructuring opportunities unexplored. We present \ours{}\unindentedfootnote{%
Our code is available at
\url{https://github.com/shuoren-mike/ReSynP3D}.\newline
\textsuperscript{*}~Corresponding author: Rongliang Fu (rlfu@cse.cuhk.edu.hk).},
a timing-driven cross-tier logic resynthesis method integrated into placement
iterations. It restructures critical combinational regions across dies using
Boolean transformations and timing-aware technology mapping under
placement-derived boundary timing and load constraints. Internal cross-tier
connections can change while the region's input--output functionality is
preserved. A measured cross-tier delay penalty guides resynthesis, and
criticality-weighted local tier repartitioning assigns the replacement cells
to the two tiers.
Together, these steps target cell delay and tier crossings, with power,
performance, and area (PPA) evaluated after routing. Across the 16 reported cases on eight
designs and two technology nodes, \ours{} improves WNS-derived post-route
frequency by 9.1\% on average and up to 34.2\% over open-source Pin-3D.
\end{abstract}

\section{Introduction}
\label{sec:intro}

As Moore's law slows, three-dimensional integrated circuits (3D ICs)
offer a promising path to continued improvements in chip performance and
energy efficiency~\cite{liu2023ieee,he2026timescaling,he2026tauchip}.
3D ICs replace long lateral connections with short
vertical ones~\cite{zhao2025survey,lu2017tsv}, and fine-pitch hybrid
bonding makes face-to-face (F2F) stacking attractive for distributing logic
across tiers~\cite{beyne2017hybrid}. At the architectural level, tier assignment
and block placement affect thermally sustainable frequency~\cite{renClipTCAD}.
At the cell level, realizing timing gains requires EDA flows in
which logic synthesis and physical design cooperate. Conventional flows run
them as separate stages~\cite{kutzschebauch2001congestion,pan2024pigmap}:
synthesis fixes a logic implementation before placement and routing (PnR)
determine cell locations and interconnect, so improvements under synthesis
estimates need not survive layout~\cite{gosti2001timing,pan2024pigmap}.
Physically aware synthesis, from layout-driven
synthesis~\cite{kutzschebauch2001congestion} to PigMAP~\cite{pan2024pigmap},
uses early placement information to guide optimization and mapping;
local resynthesis has long been interleaved with placement~\cite{saxena2004modeling}.
Placement-aware resynthesis, as in Lesyn~\cite{pu2024lesyn}, restructures
logic once cell locations exist; commercial flows such as Cadence's Critical
Region Restructuring~\cite{cadence2019crr},
iSpatial~\cite{cadence2020ispatial,cadence2020digitalflow} and Synopsys
Fusion Compiler~\cite{synopsys2018fusion} share engines and data between
synthesis and PnR.
Together, these advances highlight joint optimization of logic resynthesis
and physical design as a promising approach to improving PPA.

Coupling logic resynthesis with physical design introduces additional
challenges in 3D ICs. Unlike 2D designs, where cells occupy a single
placement plane, physical timing in 3D ICs depends on both planar cell
locations and tier assignments~\cite{zhao2025survey,pentapati2024pin3d}:
a connection may remain within one tier or cross tiers, incurring
hybrid-bond delay and the buffers that drive it~\cite{pentapati2024pin3d}.
Before 3D placement, these coupled choices are not yet resolved, making
it difficult for synthesis to identify which paths will cross tiers and
estimate their timing cost~\cite{pentapati2024pin3d}.
Iterative 3D placement provides an opportunity to address this uncertainty.
Pin-3D~\cite{pentapati2020pin3d,pentapati2024pin3d} and related pseudo-3D
flows~\cite{park2021pseudo3d} use 2D placement methods for 3D ICs.
During placement iterations, the current design exposes cell locations,
tier assignments and timing estimates, providing the physical context
needed to guide cross-tier resynthesis and local tier repartitioning.
AnchorGrid restructures logic during placement-aware 3D net-to-pad
assignment~\cite{vanna2025placement}. ResynMD resynthesizes partitioned
multi-die FPGAs before placement to reduce inter-die connections~\cite{wang2026interconnect}.
After routing, AgenticECO repairs bond-level defects in 3D ICs~\cite{ren2026agenticeco}.

\begin{figure}[t]
 \centering
 \subfloat[Before resynthesis]{\includegraphics[width=0.46\linewidth]{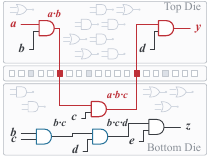}} \hfill
 \subfloat[After resynthesis]{\includegraphics[width=0.46\linewidth]{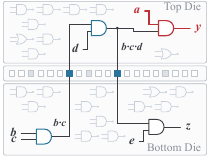}}
 \caption{Motivation for \ours{}: cross-tier resynthesis and tier
 repartitioning expose timing optimization opportunities between placement iterations.}
 \label{fig:intro}
\end{figure}

We target the interval between consecutive F2F placement iterations:
placement-derived critical paths and boundary timing guide cross-tier
resynthesis, then the replacement cells are repartitioned before placement resumes.
The current placement can reveal critical paths that repeatedly cross
tiers, motivating changes to both the logic structure and tier assignment.
\Cref{fig:intro} illustrates this opportunity. Both circuits implement
$y=abcd$ and $z=bcde$. After resynthesis and tier repartitioning, the
critical path from $a$ to $y$ passes through one gate instead of three and
no longer crosses tiers, while the remaining crossings fall on
non-critical nets (\cref{sec:prelim-resyn}).

To exploit this opportunity, we present \ours{}, a timing-aware
framework that couples cross-tier logic resynthesis with local tier
repartitioning between consecutive F2F 3D IC placement iterations to
improve post-route timing. Our contributions are summarized as follows:
\begin{itemize}[leftmargin=*,nosep]
 \item To the best of our knowledge, \ours{} is the first framework to couple
 timing-guided cross-tier logic resynthesis with local tier repartitioning
 between consecutive placement iterations for face-to-face 3D ICs.
 \item We propose placement-guided cross-tier resynthesis: critical logic
 spanning both tiers is restructured under the timing and loads of the
 current placement, with a measured crossing delay added to inputs that
 arrive from the opposite tier.
 \item To retain the resynthesis gain through placement and routing, we
 repartition only the new cells across the two tiers, keeping every other
 cell on its tier and steering tier crossings away from timing-critical nets.
 \item Experiments across multiple benchmark designs demonstrate improved
 post-route timing, with WNS-derived frequency gains of up to 34.2\%
 over an open-source Pin-3D flow.
\end{itemize}

\section{Preliminaries}
\label{sec:prelim}

\subsection{3D IC Design Flow}
\label{sec:prelim-flow}

In the 3D IC design setting studied here, we consider two device tiers
connected face-to-face through hybrid-bond
terminals. A mapped netlist $N=(V,E)$ contains cells $v\in V$ with library
area $a_v$ and nets $e\in E$; a net connects the output pin of its
\emph{driver} to the input pins of its \emph{sinks}. A tier assignment
$\pi:V\to\{0,1\}$ gives each cell's tier, and a net is cut when its cells
occupy both tiers.
At the floorplanning stage, Great3D jointly optimizes block locations and
die assignments~\cite{ren2026partitioning}, while Rect3D couples probabilistic
die assignment with rectilinear block geometry~\cite{ren2026rect3d}.
At the cell level, commercial platforms~\cite{cadenceintegrity3d} support 3D
implementation, and Pin-3D~\cite{pentapati2020pin3d,pentapati2024pin3d} adapts 2D placement
and routing engines to die assignment and cross-die connectivity.

\subsection{Logic Resynthesis and Motivating Example}
\label{sec:prelim-resyn}

Logic resynthesis changes circuit structure while preserving input--output
functionality~\cite{fu2026elogic}. In 3D ICs, its timing benefit also depends on where the
resulting cells are assigned.

As illustrated in \cref{fig:intro}, the two outputs share the intermediate
function $t=bcd$. Reusing $t$ and assigning the replacement logic to the
top die shortens the highlighted path, while the cross-tier connections
move to the shared branch rather than disappear. The example shows why
logic restructuring and tier assignment should be considered together.

\subsection{Problem Formulation}
\label{sec:problem}
\begin{table}[t]
 \centering
 \caption{Main notation.}
 \label{tab:notation}
 \footnotesize
 \renewcommand{\arraystretch}{1.12}
 \begin{tabular}{@{}p{0.235\columnwidth}@{\hspace{4pt}}p{\dimexpr0.765\columnwidth-4pt\relax}@{}}
 \toprule
 Symbol & Definition \\
 \midrule
 $N=(V,E)$ & Netlist with cells $V$ and nets $E$. \\
 $\pi, X$ & Tier assignment and planar coordinates of the cells. \\
 $\mathcal S$ & State $(N,\pi,X)$ between two placement iterations. \\
 $K, B$ & Number of worst paths selected; cell budget of the region. \\
 $R$ & Region: the editable cells on the selected paths. \\
 $E_{\mathrm{in}}, E_{\mathrm{out}}$ & Boundary nets entering and leaving the region. \\
 $A_e, t_{\mathrm{req},e}, L_e$ & Arrival time, required time, and load of net $e$. \\
 $\Theta(R)$ & Boundary conditions of the region: arrivals, driver tiers, required times, loads. \\
 $\delta_H$ & Extra delay of a tier crossing, measured on the placed design. \\
 $R'_{\rm p}, R'_{\rm u}$ & Candidates resynthesized with and without the physical boundary conditions. \\
 $R', R^*$ & A candidate replacement; the selected one. \\
 $s_e, c_e, w_e$ & Slack, criticality, and cut weight of net $e$. \\
 $\lambda, \gamma$ & Strength and exponent of the timing premium in $w_e$. \\
 $\chi_e(\pi)$ & One if net $e$ spans both tiers, zero otherwise. \\
 $H_{R'}, V_F$ & Local hypergraph of $R'$; its fixed external neighbours. \\
 $\Pi_{R'}, \pi_{R'}$ & Feasible tier assignments of $R'$; the one chosen. \\
 $\mathcal Q$ & Candidates passing the equivalence and feasibility checks. \\
 \bottomrule
 \end{tabular}
\end{table}

The main notation is summarized in \Cref{tab:notation}.
Motivated by \cref{fig:intro}, our objective is to improve the timing
of a 3D IC by resynthesizing critical logic and reassigning the resulting
cells to the two dies. Given the current placement state $\mathcal S=(N,\pi,X)$ with
netlist $N$, tier assignment $\pi$ and cell coordinates $X$, we seek a functionally equivalent
netlist $N'$ and a new assignment $\pi'$ that maximize the post-route worst
setup slack in \cref{eq:problem}:
\begin{equation}
 \max_{N',\pi'}\quad \mathrm{WNS}_{\mathrm{route}}(N',\pi').
 \label{eq:problem}
\end{equation}
Here $\mathrm{WNS}_{\mathrm{route}}$ is the worst register-to-register setup slack of the
routed design with extracted parasitics.

\section{Methodology}
\label{sec:method}

\begin{figure}[t]
 \centering
 \begingroup
\definecolor{flowblue}{HTML}{426E98}
\definecolor{flowborder}{HTML}{89939C}
\definecolor{flowgray}{HTML}{E4E6E8}
\definecolor{flowgreen}{HTML}{E6F0EB}
\definecolor{flowgold}{HTML}{FFF2D8}
\definecolor{flowpurple}{HTML}{EEE8F5}
\begin{tikzpicture}[x=1mm,y=1mm,
  font=\fontsize{7}{8}\selectfont,
  box/.style={draw=flowborder,line width=.4pt,rounded corners=.4mm,align=center,inner sep=1mm},
  arrow/.style={-{Latex[length=1.2mm,width=.9mm]},draw=flowblue,line width=.55pt},
  title/.style={font=\fontsize{7}{8}\selectfont\bfseries,align=center,inner sep=0pt},
  cell/.style={box,fill=white,fill opacity=.65,text opacity=1,minimum width=22.1mm,minimum height=8mm,text width=21.5mm,inner sep=.3mm},
  side/.style={font=\fontsize{7}{8}\selectfont\bfseries,text=flowblue,rotate=90,inner sep=0pt}]
\node[box,fill=flowgray,minimum width=82mm,minimum height=5mm,text=flowblue,font=\fontsize{7}{8}\selectfont\bfseries] (input) at (41,61.5) {Input: placed netlist with tier assignment};

\fill[flowgreen,rounded corners=.6mm] (0,42) rectangle (9,57);
\draw[box,fill=flowgreen] (11.5,42) rectangle (82,57);
\node[side] at (4.5,49.5) {Extraction};
\node[title] at (46.75,54.2) {A. Timing-guided region extraction (\cref{sec:extraction})};
\node[cell] at (23.9,47.7) {$K$ worst paths\\by slack};
\node[cell] at (46.75,47.7) {Freeze the\\region boundary};
\node[cell] at (69.6,47.7) {Expand selected\\cell functions};

\fill[flowgold,rounded corners=.6mm] (0,25) rectangle (9,40);
\draw[box,fill=flowgold] (11.5,25) rectangle (82,40);
\node[side] at (4.5,32.5) {Resynthesis};
\node[title] at (46.75,37.2) {B. Cross-tier logic resynthesis (\cref{sec:resynthesis})};
\node[cell] at (23.9,30.7) {Restructure\\the logic};
\node[cell] at (46.75,30.7) {Map with boundary\\timing, crossing cost};
\node[cell] at (69.6,30.7) {Two mapped\\candidates};

\fill[flowpurple,rounded corners=.6mm] (0,8) rectangle (9,23);
\draw[box,fill=flowpurple] (11.5,8) rectangle (82,23);
\node[side] at (4.5,15.5) {Partitioning};
\node[title] at (46.75,20.2) {C. Timing-aware tier repartitioning (\cref{sec:repartition})};
\node[cell] at (23.9,13.7) {Score each net\\by criticality};
\node[cell] at (46.75,13.7) {Partition new cells,\\keep old tiers};
\node[cell] at (69.6,13.7) {Full-design STA\\selects a candidate};

\node[box,fill=flowgray,minimum width=25.5mm,minimum height=7mm,inner sep=.5mm] (apply) at (12.75,3.5) {Apply the update};
\node[box,fill=flowgray,minimum width=25.5mm,minimum height=7mm,inner sep=.5mm] (follow) at (41,3.5) {Following physical\\design flow};
\node[box,fill=flowgray,minimum width=25.5mm,minimum height=7mm,inner sep=.5mm,text=flowblue,font=\fontsize{7}{8}\selectfont\bfseries] (out) at (69.25,3.5) {Post-route PPA};
\draw[arrow] (apply.east) -- (follow.west);
\draw[arrow] (follow.east) -- (out.west);
\foreach \top/\bottom in {59/57,42/40,25/23,8/7} {
  \draw[arrow] (46.75,\top) -- (46.75,\bottom);
  \draw[arrow] (4.5,\top) -- (4.5,\bottom);
}
\foreach \mid/\top/\bottom in {49.5/56/43,32.5/39/26,15.5/22/9} {
  \draw[flowblue,dashed,line width=.4pt] (11.2,\top) -- (9.3,\mid) -- (11.2,\bottom);
}
\end{tikzpicture}
\endgroup
 \caption{ReSynP3D workflow: region extraction, cross-tier resynthesis,
 and timing-aware tier repartitioning. Each stage supplies the next.}
 \label{fig:method-flow}
\end{figure}
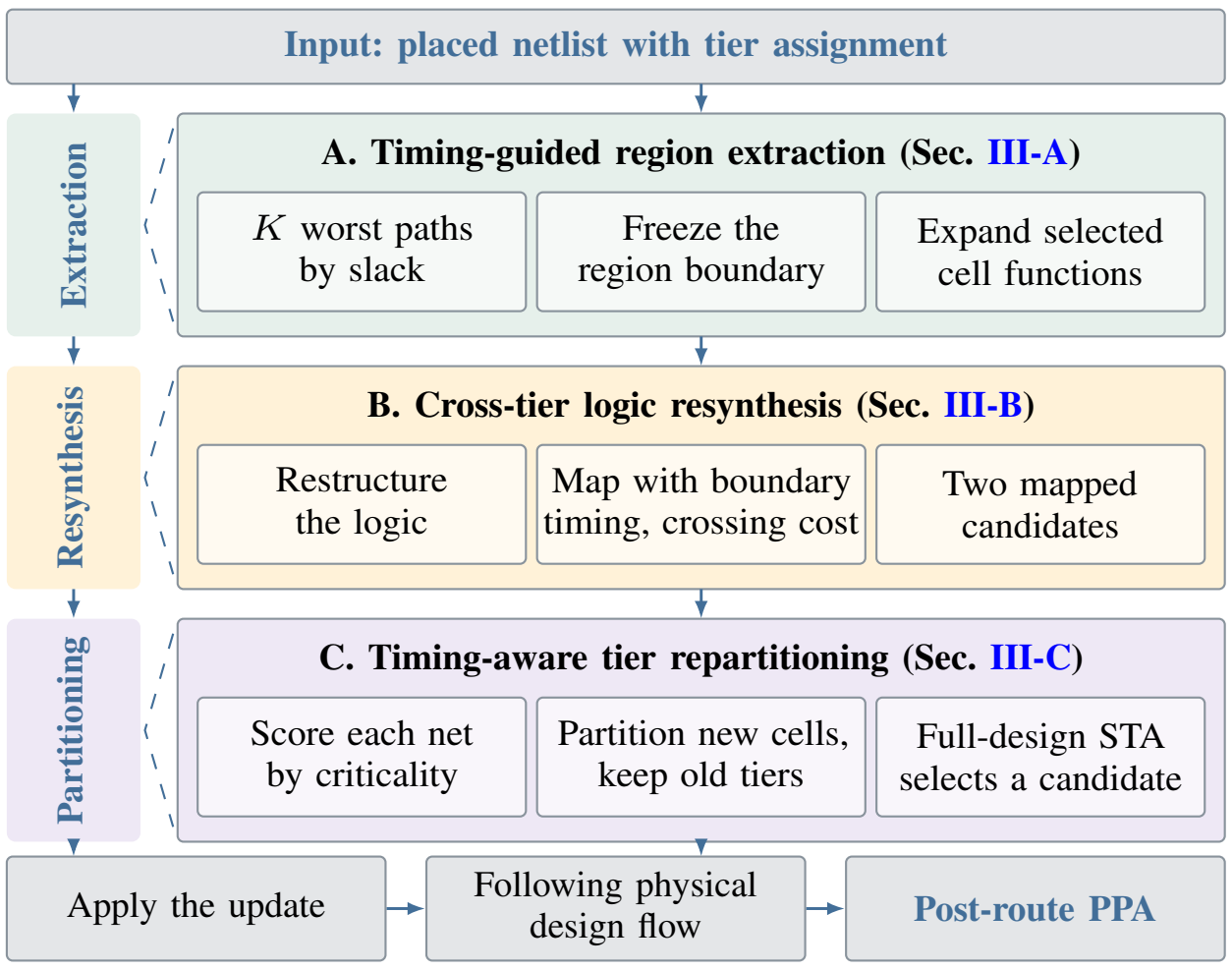

To solve the problem formulated in \cref{sec:problem}, we propose
the \ours{} framework shown in \cref{fig:method-flow}.

\subsection{Timing-Guided Region Extraction}
\label{sec:extraction}

\suppressfloats[t]

In Pin-3D-based flows~\cite{pentapati2024pin3d,jiang2026benchmarking},
each placement iteration optimizes one tier while the other is held fixed
and then swaps their roles. \ours{} operates in the gap between two
such iterations, where every cell already has a tier and a location and
timing is available as a placement-stage estimate. We use this
intermediate state to prepare the input of the cross-tier resynthesis in
\cref{sec:resynthesis}: the logic to rebuild and the constraints
it must meet.

Placement-stage STA ranks paths by estimated slack, using library cell
delays and wire delays estimated from net length with uniform per-unit RC;
it does not include an explicit tier-crossing penalty. We take the $K$
worst paths and merge their editable combinational cells into one region
$R$. Shared cells are included once, allowing overlapping paths to be
optimized jointly. If the region exceeds the cell budget $B$, we reduce
the selected path count while retaining the worst paths first. We report
$B$ as a fraction of the current design's cell count in the experiments.

The region may span both tiers. We preserve its connections to the
surrounding circuit while allowing its internal logic, including cross-tier
connections, to change. Its interface therefore follows connectivity: the
input nets $E_{\mathrm{in}}$ are driven outside the region and feed cells
inside, and the output nets $E_{\mathrm{out}}$ are driven inside and feed
cells outside.

To guide resynthesis, we record when inputs arrive, when outputs are
required and what loads they drive. Input-driver tiers inform
the next stage's crossing penalty. We collect these boundary conditions as
\begin{equation}
 \Theta(R)=\{(A_e,\pi(u_e))\}_{e\in E_{\mathrm{in}}}
 \cup\{(t_{\mathrm{req},e},L_e)\}_{e\in E_{\mathrm{out}}},
 \label{eq:boundary-env}
\end{equation}
where $A_e$ is the arrival time of input net $e$, $u_e$ its driver, and
$t_{\mathrm{req},e}$ the required time of output net $e$. The load
$L_e$ is the capacitance of output net $e$ seen by its driver. The boundary nets
become the primary inputs and outputs of the resynthesis problem.
After recording these constraints, we expand the selected cells into
Boolean logic for resynthesis, preserving the region's interface.

\subsection{Cross-Tier Logic Resynthesis}
\label{sec:resynthesis}

Given the region and its boundary conditions, resynthesis seeks a faster
implementation of the selected logic while preserving its boundary
functions. It can change logic depth, sharing and cell choices, including
using more cells when this reduces delay.

Without physical context, synthesis must assume common arrival, drive and
load conditions at the region boundary. Here the region sits in a placed
3D design, and the boundary conditions $\Theta(R)$ of
\cref{eq:boundary-env} describe its surroundings, so late inputs, tight
output requirements and output loads guide the new implementation.
Tier crossings add a further effect. For an input whose driver is on the
opposite tier from the majority of the sinks of the region's outputs, the
arrival that the mapper sees also carries a crossing penalty:
\begin{equation}
 A_e\leftarrow A_e+\delta_H\,\mathbf 1[\pi(u_e)\ne t_R],
 \label{eq:boundary-arrival}
\end{equation}
where $t_R$ is that majority tier and $\delta_H$ the estimated crossing
penalty. We estimate $\delta_H$ from the placed design using
cross-tier and same-tier nets sampled with matched fanout:
\begin{equation}
 \delta_H=\operatorname*{med}_{e\in E_\times}\bigl(A^{\rm snk}_e-A^{\rm drv}_e\bigr)
         -\operatorname*{med}_{e\in E_=}\bigl(A^{\rm snk}_e-A^{\rm drv}_e\bigr),
 \label{eq:hbt-penalty}
\end{equation}
where $E_\times$ and $E_=$ are the two net sets, and $A^{\rm snk}_e$ and
$A^{\rm drv}_e$ are the latest sink and driver arrivals.
This penalty guides synthesis; it does not assign the new cells to tiers.

Under these conditions, we restructure the extracted AND-inverter graph
with passes such as balancing, rewriting and refactoring, and technology-map it back to
standard cells with buffering and sizing. The mapper targets the worst
boundary slack of the replacement $R'$:
\begin{equation}
 \min_{R'}\ \max_{o\in E_{\mathrm{out}}}\bigl(A_o-t_{\mathrm{req},o}\bigr),
 \label{eq:map-target}
\end{equation}
where $A_o$ is the arrival time at output $o$ of $R'$, propagated through
library delays from the input arrivals. The mapper approaches this
objective heuristically. This local delay objective serves the final
post-route WNS goal. Synthesis omits internal wire delays because
subsequent placement relocates the new cells. The new logic must
implement the same boundary functions as the original region;
equivalence with the original region is checked at every boundary output.

Because the mapper is heuristic, physical constraints do not guarantee a
better implementation. As a safeguard, we also resynthesize the
region under the common boundary assumptions. The two results, the physical
candidate $R'_{\rm p}$ and the uniform candidate $R'_{\rm u}$, both
proceed to tier repartitioning, where one of them is selected.

\subsection{Timing-Aware Tier Repartitioning}
\label{sec:repartition}
Resynthesis changes both the cells and their connections, so the new
logic needs a tier assignment suited to its timing. We repartition each
mapped candidate separately while keeping external cells fixed. Critical nets receive higher cut costs,
encouraging the partitioner to avoid crossings on timing-sensitive
connections.
For each candidate $R'$, STA on its mapped cells uses the extracted
boundary arrivals, required times and loads, with ideal internal wires.
This scores the new logic without requiring an initial tier assignment
or placement. On the local hypergraph $H_{R'}=(R'\cup V_F,E_{R'})$,
$V_F$ contains fixed external neighbours and $E_{R'}$ the nets incident to
candidate cells. Let $s_e$ be the minimum pin slack on net $e$ and
$E_{\rm STA}$ the nets with available timing. Their normalized criticality is
\begin{equation}
 c_e=\frac{s_{\max}-s_e}{s_{\max}-s_{\min}},
 \qquad e\in E_{\rm STA},
 \label{eq:net-criticality}
\end{equation}
\begin{algorithm}[t]
 \caption{\ours{} between placement iterations}
 \label{alg:resynp3d}
 \small
 \DontPrintSemicolon
 \SetCommentSty{textnormal}
 \SetKwInput{KwIn}{Input}
 \SetKwInput{KwOut}{Output}
 \KwIn{State $\mathcal S=(N,\pi,X)$}
 \KwOut{State $\mathcal S^+$ after placement continuation}
 $\mathcal S'\gets\mathcal S$\nllabel{ln:retain}\;
 $R\gets\textsc{Extract}(\mathcal S,K,B)$\nllabel{ln:extract} \tcp*{\cref{sec:extraction}}
 $\{R'_{\rm u},R'_{\rm p}\}\gets\textsc{Resynthesize}(\mathcal S,R)$\nllabel{ln:resynthesize}\;
 $\mathcal Q\gets\varnothing$\;
 \ForEach{valid candidate $R'$\nllabel{ln:valid}}{
  $H_{R'}\gets\textsc{Hypergraph}(R',\mathcal S)$\nllabel{ln:hypergraph}\;
  $T'\gets\textsc{BoundarySTA}(R',\Theta(R))$\;
  \ForEach{net $e\in E_{R'}$}{
   $s_e\gets\textsc{NetSlack}(T',e)$\nllabel{ln:slack}\;
   $c_e\gets\textsc{Criticality}(s_e,T')$\nllabel{ln:criticality} \tcp*{\cref{eq:net-criticality}}
   $w_e\gets 1+\lambda c_e^{\gamma}$\;
  }
  $\Pi_{R'}\gets\textsc{FeasibleSet}(R',\pi)$\nllabel{ln:feasible-set}\;
  $\pi_{R'}\gets\textsc{TierPartition}(H_{R'},\mathbf w,\Pi_{R'})$\nllabel{ln:reassign}\;
  \If{$\mathrm{Feasible}(\pi_{R'})$\nllabel{ln:feasible}}{
   $\textsc{Instantiate}(\mathcal S,R',\pi_{R'})$\;
   $\mathcal Q\gets\mathcal Q\cup\{R'\}$\;
  }
 }
 \If{$\mathcal Q\ne\varnothing$}{
  $R^*\gets\textsc{SelectBySTA}(\mathcal Q)$ \tcp*{\cref{sec:repartition}}
  $\mathcal S'\gets\textsc{ApplyReplacement}(\mathcal S,R^*,\pi_{R^*})$\nllabel{ln:apply}\;
 }
 $\mathcal S^+\gets\textsc{ContinuePlacementIteration}(\mathcal S')$\nllabel{ln:continue}\;
 \KwRet{$\mathcal S^+$}\;
\end{algorithm}

where $s_{\min}$ and $s_{\max}$ are the slack extremes over $E_{\rm STA}$.
We assume a nonzero slack range, $s_{\max}>s_{\min}$.
Nets without timing data receive $c_e=0$. The most critical scored net
receives one and the most relaxed receives zero.
The cut weight $w_e=1+\lambda c_e^{\gamma}$ combines a unit cost for
every cut with a timing premium: $\lambda$ controls its strength and
$\gamma\ge1$ concentrates it on the critical nets. The weights form
the vector $\mathbf w$ used in \cref{alg:resynp3d}.
The partitioning objective is
\begin{equation}
 \pi_{R'}=\arg\min_{\pi'\in\Pi_{R'}}\sum_{e\in E_{R'}}w_e\,\chi_e(\pi'),
 \label{eq:local-tier-cut}
\end{equation}
where $\chi_e(\pi')$ is one if net $e$ spans both tiers and zero
otherwise. The feasible set $\Pi_{R'}$ contains the assignments that keep
every external cell on its current tier. The replacement is small
relative to the two tiers, which the flow has already balanced, so the
area budget of each tier is sized to admit the new cells and the weighted cut
decides their split; die capacity and placement legality are left to the
placement that follows.
Hypergraph partitioning approximately solves this weighted-cut problem
with external neighbours pinned.

After repartitioning, the new cells of each candidate are bound to the
library cells of their assigned tiers and inserted into a copy of the
current design, each starting near the centroid of its fixed neighbours. A candidate is kept only if it is functionally equivalent
and its tier assignment is feasible. Among the candidates kept,
full-design placement-stage STA, which sees the new tiers and starting
locations, selects the one with the better register-to-register WNS. The selected candidate is applied and placement
resumes to refine cell locations; no further technology mapping is needed
at this handoff. The final timing benefit is evaluated after routing, and
no routed timing is fed back into the update.

\stepcounter{table}
\begin{table*}[t]
\caption{Placement-stage WNS, post-route PPA and matched-endpoint timing for eight designs on two nodes,
Pin-3D versus \ours{}.}
\label{tab:main}
\centering
\footnotesize
\newsavebox{\mainresultsbox}
\setlength{\tabcolsep}{2pt}
\sbox{\mainresultsbox}{%
\begin{tabular}{llr*{21}{r}}
\toprule
\multirow{2.5}{*}{Design} & \multirow{2.5}{*}{Node} & \multirow{2.5}{*}{$T_{\mathrm{clk}}$} & \multicolumn{8}{c}{Pin-3D} & \multicolumn{8}{c}{\ours{}} & \multicolumn{5}{c}{Matched endpoints} \\
\cmidrule(lr){4-11}\cmidrule(lr){12-19}\cmidrule(lr){20-24}
 & & & $\mathrm{WNS}_{\mathrm{place}}\uparrow$ & $\mathrm{WNS}_{\mathrm{route}}\uparrow$ & TNS$\uparrow$ & $F\uparrow$ & Area$\downarrow$ & Power$\downarrow$ & WL$\downarrow$ & HBT$\downarrow$ & $\mathrm{WNS}_{\mathrm{place}}\uparrow$ & $\mathrm{WNS}_{\mathrm{route}}\uparrow$ & TNS$\uparrow$ & $F\uparrow$ & Area$\downarrow$ & Power$\downarrow$ & WL$\downarrow$ & HBT$\downarrow$ & $N_{\mathrm m}$ & Better & $\Delta$Slack & $\Delta$Cell & $\Delta$Buf \\
\midrule
\multirow{2}{*}{JPEG} & ASAP7 & 620 & $-0.404$ & $-0.950$ & $-371$ & $0.64$ & $6.71$ & $135.0$ & $162.4$ & $1397$ & $\mathbf{-0.384}$ & $\mathbf{-0.550}$ & $\mathbf{-230}$ & $\mathbf{0.85}$ & $6.94$ & $136.0$ & $170.4$ & $1498$ & $239$ & $233$ & $286$ & $-179$ & $-59$ \\
 & Nangate45 & 1000 & $-0.609$ & $-1.849$ & $-2167$ & $0.35$ & $94.51$ & $483.0$ & $595.1$ & $1568$ & $-0.630$ & $\mathbf{-1.505}$ & $\mathbf{-1113}$ & $\mathbf{0.40}$ & $97.21$ & $\mathbf{472.0}$ & $635.2$ & $1634$ & $256$ & $254$ & $417$ & $-20$ & $162$ \\
\midrule
\multirow{2}{*}{SHA-3} & ASAP7 & 600 & $-0.341$ & $-0.620$ & $-521$ & $0.82$ & $8.15$ & $921.0$ & $488.3$ & $4685$ & $-0.375$ & $\mathbf{-0.557}$ & $-616$ & $\mathbf{0.86}$ & $8.27$ & $951.0$ & $496.7$ & $4769$ & $267$ & $101$ & $-17$ & $10$ & $19$ \\
 & Nangate45 & 1500 & $-0.074$ & $-0.578$ & $-631$ & $0.48$ & $65.46$ & $744.0$ & $1077.3$ & $4404$ & $-0.076$ & $\mathbf{-0.443}$ & $\mathbf{-467}$ & $\mathbf{0.51}$ & $73.48$ & $852.0$ & $1119.0$ & $4655$ & $197$ & $196$ & $107$ & $-242$ & $138$ \\
\midrule
\multirow{2}{*}{PicoRV32} & ASAP7 & 700 & $0.084$ & $-0.122$ & $-7$ & $1.22$ & $2.29$ & $17.6$ & $86.0$ & $832$ & $\mathbf{0.096}$ & $\mathbf{-0.026}$ & $\mathbf{-0}$ & $\mathbf{1.38}$ & $\mathbf{2.28}$ & $\mathbf{17.4}$ & $88.1$ & $836$ & $300$ & $160$ & $48$ & $24$ & $-66$ \\
 & Nangate45 & 700 & $-0.138$ & $-0.305$ & $-121$ & $1.00$ & $31.41$ & $76.4$ & $318.1$ & $1227$ & $-0.177$ & $\mathbf{-0.278}$ & $\mathbf{-120}$ & $\mathbf{1.02}$ & $32.10$ & $78.4$ & $328.7$ & $1255$ & $456$ & $238$ & $4$ & $21$ & $-22$ \\
\midrule
\multirow{2}{*}{SweRV} & ASAP7 & 1360 & $0.048$ & $-0.037$ & $-0$ & $0.72$ & $12.85$ & $138.0$ & $948.8$ & $1597$ & $\mathbf{0.110}$ & $\mathbf{0.065}$ & $\mathbf{0}$ & $\mathbf{0.77}$ & $12.86$ & $139.0$ & $\mathbf{947.6}$ & $1597$ & $350$ & $289$ & $29$ & $-155$ & $127$ \\
 & Nangate45 & 1900 & $-0.186$ & $-0.747$ & $-1214$ & $0.38$ & $163.18$ & $448.0$ & $3473.7$ & $2166$ & $\mathbf{-0.176}$ & $\mathbf{-0.734}$ & $-1272$ & $\mathbf{0.38}$ & $164.04$ & $497.0$ & $3520.8$ & $2186$ & $433$ & $161$ & $-13$ & $30$ & $-3$ \\
\midrule
\multirow{2}{*}{SHA-256} & ASAP7 & 500 & $-0.326$ & $-0.613$ & $-134$ & $0.90$ & $2.65$ & $204.0$ & $97.3$ & $1241$ & $-0.372$ & $\mathbf{-0.493}$ & $\mathbf{-120}$ & $\mathbf{1.01}$ & $2.73$ & $\mathbf{202.0}$ & $100.2$ & $1426$ & $491$ & $286$ & $30$ & $12$ & $-15$ \\
 & Nangate45 & 1100 & $-0.283$ & $-0.495$ & $-110$ & $0.63$ & $36.98$ & $555.0$ & $335.3$ & $1371$ & $-0.300$ & $\mathbf{-0.463}$ & $-116$ & $\mathbf{0.64}$ & $37.83$ & $575.0$ & $340.7$ & $1382$ & $495$ & $200$ & $-11$ & $-30$ & $48$ \\
\midrule
\multirow{2}{*}{SystemC-AES} & ASAP7 & 600 & $0.012$ & $-0.007$ & $-0$ & $1.65$ & $0.88$ & $12.3$ & $36.9$ & $806$ & $0.007$ & $\mathbf{-0.003}$ & $\mathbf{-0}$ & $\mathbf{1.66}$ & $0.89$ & $12.9$ & $37.5$ & $\mathbf{805}$ & $494$ & $315$ & $8$ & $-29$ & $21$ \\
 & Nangate45 & 1100 & $0.053$ & $-0.435$ & $-45$ & $0.65$ & $10.47$ & $29.7$ & $115.6$ & $782$ & $-0.043$ & $\mathbf{-0.090}$ & $\mathbf{-0}$ & $\mathbf{0.84}$ & $10.89$ & $32.6$ & $119.7$ & $790$ & $483$ & $230$ & $71$ & $-192$ & $121$ \\
\midrule
\multirow{2}{*}{SHA-core} & ASAP7 & 600 & $-0.071$ & $-0.093$ & $-2$ & $1.44$ & $0.94$ & $18.1$ & $28.9$ & $214$ & $-0.081$ & $\mathbf{-0.078}$ & $-3$ & $\mathbf{1.47}$ & $0.95$ & $\mathbf{18.0}$ & $29.2$ & $218$ & $350$ & $151$ & $-15$ & $-10$ & $27$ \\
 & Nangate45 & 1100 & $-0.079$ & $-0.191$ & $-3$ & $0.77$ & $12.00$ & $49.7$ & $106.8$ & $333$ & $-0.114$ & $\mathbf{-0.160}$ & $-3$ & $\mathbf{0.79}$ & $12.38$ & $55.6$ & $107.0$ & $333$ & $415$ & $214$ & $1$ & $-44$ & $61$ \\
\midrule
\multirow{2}{*}{DynamicNode} & ASAP7 & 700 & $-0.166$ & $-0.217$ & $-5$ & $1.09$ & $1.87$ & $14.9$ & $66.2$ & $1028$ & $-0.205$ & $\mathbf{-0.178}$ & $\mathbf{-4}$ & $\mathbf{1.14}$ & $1.87$ & $15.0$ & $66.3$ & $1029$ & $258$ & $194$ & $18$ & $-21$ & $4$ \\
 & Nangate45 & 1080 & $-0.019$ & $-0.260$ & $-8$ & $0.75$ & $23.47$ & $38.4$ & $223.4$ & $1215$ & $-0.053$ & $\mathbf{-0.153}$ & $\mathbf{-5}$ & $\mathbf{0.81}$ & $23.68$ & $38.8$ & $227.0$ & $1215$ & $367$ & $119$ & $0$ & $-12$ & $7$ \\
\bottomrule
\end{tabular}}
\begin{threeparttable}
\resizebox{\linewidth}{!}{\usebox{\mainresultsbox}}
\begin{tablenotes}[flushleft]
\footnotesize
\item[*] $T_{\mathrm{clk}}$: target clock period, ps; WNS/TNS: ns; $F$: effective frequency, GHz; Area $10^3\,\mu\mathrm{m}^2$; Power mW; WL mm; bold is better. Better counts improved endpoints; $\Delta$Slack, $\Delta$Cell (cell), $\Delta$Buf (buffer, crossing) = \ours{} $-$ Pin-3D, mean ps.
\end{tablenotes}
\end{threeparttable}
\end{table*}

\suppressfloats[t]
\setcounter{topnumber}{1}
\addtocounter{table}{-2}
\begin{table}[t]
\caption{Benchmark statistics in our experiments.}
\label{tab:bench}
\centering
\experimenttablefont
\begin{tabular*}{\columnwidth}{@{\extracolsep{\fill}}l*{10}{r}@{}}
\toprule
 & \multicolumn{2}{c}{} & \multicolumn{4}{c}{ASAP7} & \multicolumn{4}{c}{Nangate45} \\
\cmidrule(lr){2-3}\cmidrule(lr){4-7}\cmidrule(lr){8-11}
Design & \#PI & \#PO & \#Cell & \#Net & \#Reg & $T_{\mathrm{clk}}$ & \#Cell & \#Net & \#Reg & $T_{\mathrm{clk}}$ \\
\midrule
 JPEG & 20 & 27 & 52916 & 60835 & 4383 & 620 & 57408 & 69710 & 4383 & 1000 \\
 SHA-3 & 71 & 514 & 50127 & 50198 & 2211 & 600 & 56774 & 59056 & 2211 & 1500 \\
 PicoRV32 & 67 & 173 & 17985 & 18419 & 2367 & 700 & 18892 & 21693 & 2367 & 700 \\
 SweRV & 534 & 882 & 84322 & 88061 & 11256 & 1360 & 79586 & 94632 & 11306 & 1900 \\
 SHA-256 & 514 & 256 & 17284 & 19538 & 2304 & 500 & 19780 & 24338 & 2304 & 1100 \\
 SystemC-AES & 260 & 129 & 6470 & 6752 & 674 & 600 & 6682 & 7638 & 674 & 1100 \\
 SHA-core & 38 & 36 & 6624 & 6988 & 893 & 600 & 7465 & 8722 & 893 & 1100 \\
 DynamicNode & 362 & 331 & 14370 & 14897 & 2257 & 700 & 12679 & 15463 & 2257 & 1080 \\
\bottomrule
\end{tabular*}
\end{table}

\stepcounter{table}
\begin{table}[t]
\caption{Runtime breakdown (s) and \ours{} share.}
\label{tab:runtime}
\centering
\experimenttablefont
\begin{tabular*}{\columnwidth}{@{\extracolsep{\fill}}l*{10}{c}@{}}
\toprule
 & \multicolumn{5}{c}{ASAP7} & \multicolumn{5}{c}{Nangate45} \\
\cmidrule(lr){2-6}\cmidrule(lr){7-11}
Design & \ours{} & Place & Route & Flow & Ratio (\%) & \ours{} & Place & Route & Flow & Ratio (\%) \\
\midrule
 JPEG & 140 & 4450 & 1602 & 10558 & 1.3 & 138 & 2441 & 1878 & 5794 & 2.4 \\
 SHA-3 & 143 & 2663 & 2320 & 7232 & 2.0 & 119 & 4874 & 2467 & 9524 & 1.2 \\
 PicoRV32 & 69 & 315 & 730 & 1561 & 4.4 & 60 & 391 & 715 & 1459 & 4.1 \\
 SweRV & 152 & 1374 & 3300 & 5896 & 2.6 & 112 & 1662 & 2574 & 5188 & 2.2 \\
 SHA-256 & 88 & 1267 & 888 & 3783 & 2.3 & 67 & 332 & 685 & 1434 & 4.7 \\
 SystemC-AES & 54 & 106 & 325 & 663 & 8.1 & 35 & 82 & 253 & 453 & 7.7 \\
 SHA-core & 55 & 95 & 254 & 592 & 9.3 & 34 & 91 & 182 & 404 & 8.4 \\
 DynamicNode & 59 & 171 & 635 & 1236 & 4.8 & 42 & 158 & 424 & 809 & 5.2 \\
\bottomrule
\end{tabular*}
\end{table}

\begin{table}[!t]
\caption{Placement-stage selection against post-route timing.}
\label{tab:overlap}
\centering
\experimenttablefont
\begin{tabular*}{\columnwidth}{@{\extracolsep{\fill}}l*{16}{r}@{}}
\toprule
 & \multicolumn{2}{c}{JPEG} & \multicolumn{2}{c}{SHA-3} & \multicolumn{2}{c}{PicoRV32} & \multicolumn{2}{c}{SweRV} & \multicolumn{2}{c}{SHA-256} & \multicolumn{2}{c}{{\fontsize{5.5}{6.6}\selectfont SystemC-AES}} & \multicolumn{2}{c}{SHA-core} & \multicolumn{2}{c}{{\fontsize{5.5}{6.6}\selectfont DynamicNode}} \\
\cmidrule(lr){2-3}\cmidrule(lr){4-5}\cmidrule(lr){6-7}\cmidrule(lr){8-9}\cmidrule(lr){10-11}\cmidrule(lr){12-13}\cmidrule(lr){14-15}\cmidrule(lr){16-17}
 & A7 & N45 & A7 & N45 & A7 & N45 & A7 & N45 & A7 & N45 & A7 & N45 & A7 & N45 & A7 & N45 \\
\midrule
 $K$ & 408 & 386 & 500 & 332 & 247 & 204 & 500 & 500 & 95 & 81 & 77 & 75 & 48 & 26 & 229 & 183 \\
 Kept & 273 & 244 & 200 & 97 & 227 & 201 & 316 & 6 & 95 & 81 & 77 & 75 & 48 & 26 & 183 & 103 \\
 Top-100 & 100 & 49 & 47 & 19 & 100 & 70 & 59 & 0 & 36 & 60 & 65 & 47 & 29 & 26 & 96 & 98 \\
 Paths-base & 323 & 346 & 241 & 208 & 277 & 256 & 444 & 25 & 275 & 284 & 480 & 325 & 498 & 28 & 347 & 320 \\
 Paths-ours & 343 & 443 & 279 & 363 & 225 & 225 & 459 & 2 & 287 & 317 & 385 & 435 & 366 & 28 & 232 & 273 \\
 Original & 2622 & 2842 & 3055 & 2808 & 897 & 896 & 924 & 747 & 999 & 1178 & 333 & 316 & 360 & 367 & 689 & 652 \\
 Resyn & 3703 & 4163 & 3304 & 2887 & 991 & 1053 & 1039 & 1044 & 1280 & 1485 & 375 & 342 & 458 & 438 & 778 & 708 \\
\bottomrule
\end{tabular*}
\end{table}

\subsection{Overall Algorithm}
\label{sec:overall-algorithm}

\Cref{alg:resynp3d} combines the three stages into one update between
consecutive placement iterations. Line~\ref{ln:retain} retains the original
state as a fallback. Lines~\ref{ln:extract}--\ref{ln:resynthesize} extract
the critical region and its boundary conditions, then generate two
functionally equivalent, technology-mapped candidates.

Each candidate is scored by regional STA and repartitioned with external
cells fixed (lines~\ref{ln:valid}--\ref{ln:reassign}). The feasible
candidates are instantiated with their assigned tiers and initial
locations, collected in $\mathcal Q$, then compared by full-design STA
(lines~\ref{ln:feasible}--\ref{ln:apply}). Placement continues from the
selected update; all checks passed in the reported cases.

The update changes only the selected region: its boundary functions are
preserved, and every cell outside it keeps its tier and location. Its
cost is the timing analysis, resynthesis and partitioning of the two
candidates, which stays below 10\% of the flow runtime (\cref{tab:runtime}).
The final timing benefit is evaluated after routing (\cref{sec:exp}).

\section{Experimental Evaluation}
\label{sec:exp}

\subsection{Experiment Setup}
Experiments run on dual Intel Xeon Gold 6426Y processors with 256 GB RAM.
The baseline, denoted Pin-3D, is the open-source Pin-3D~\cite{pentapati2020pin3d}
flow of TaiWei~\cite{jiang2026benchmarking}, which runs placement, clock-tree
synthesis, routing and post-route timing in OpenROAD~\cite{ajayi2019openroad}.
The input netlists and the expansion of cell functions into Boolean logic use
Yosys~\cite{wolf2013yosys}; restructuring and mapping use
ABC~\cite{brayton2010abc}.
Our implementation is publicly available.\footnote{%
The repository (\url{https://github.com/shuoren-mike/ReSynP3D}) provides the implementation in resynp3d/, the Docker
build environment in Dockerfile, method parameters and run
configurations in configs/, execution scripts in run/, and the inputs of all cases in inputs/.}
The main evaluation uses eight designs: JPEG~\cite{design_jpeg}, SHA-3,
PicoRV32, SweRV~\cite{swerv_eh1}, SHA-256, SystemC-AES, SHA-core and
DynamicNode~\cite{tziantzioulis2021opdb},
each implemented with the ASAP7 7-nm PDK~\cite{clark2016asap7} and the Nangate45 Open Cell Library (45 nm)~\cite{nangate45library}.
\Cref{tab:bench} gives the size of each case as the evaluation flow receives it.

\subsection{\ours{} Timing Improvement}
We compare this flow with and without the \ours{} update, from the same
released inputs and initial tier assignment. As in AgenticPD~\cite{ren2027agenticpd},
we evaluate timing after routing. \Cref{tab:main} reports timing, power, area, wirelength and
HBT count.

\ours{} improves register-to-register WNS in all 16 reported cases, by 105 ps on
ASAP7 and 129 ps on Nangate45 on average; the largest gains are JPEG/ASAP7
(400 ps), SystemC-AES/Nangate45 (345 ps) and JPEG/Nangate45 (344 ps). The
WNS gains are accompanied by TNS tradeoffs in some cases. The largest
power and area increases occur for SHA-3/Nangate45, at approximately
15\% and 12\%, respectively, alongside a 135\,ps WNS improvement.
Elsewhere, area changes remain within about 4\%, wirelength within
about 7\%, and HBT count changes by at most 185 nets.

In the effective frequency $F=1/(T_{\mathrm{clk}}-\mathrm{WNS}_{\mathrm{route}})$
of TP-GNN~\cite{lu2020tpgnn}, the gain is 9.1\% on average over the 16 cases
and 34.2\% at most, on JPEG/ASAP7.

The matched-endpoint results help explain the timing gains.
Here $N_{\mathrm m}$ counts the endpoints both flows rank among their
500 worst (1000 for SweRV). For JPEG/ASAP7, mean cell and buffer/crossing delays decrease
by approximately 179\,ps and 59\,ps, respectively, and 233 of 239 matched
endpoints improve. Across 10 of the 16 cases, more than half of the
matched endpoints are faster. These results suggest that the update
helps improve timing across a broader set of paths, beyond the worst
endpoint alone. SweRV/Nangate45 and SHA-core/Nangate45 also show WNS
gains, although only 2 and 28 of their worst paths contain resynthesized
cells; direct attribution to logic restructuring is therefore limited
in these two cases.

The $\mathrm{WNS}_{\mathrm{place}}$ column reports register-to-register
WNS after the final detailed-placement pass, using estimated parasitics,
a later stage than the candidate comparison in
\cref{sec:repartition}.

Beyond the WNS gains, \ours{} retains competitive power and area:
both remain within 5\% of Pin-3D in 12 of the 16 reported cases
(\cref{tab:main}), with larger tradeoffs in the remaining Nangate45
cases. \Cref{tab:runtime} reports the time spent in \ours{},
placement and routing alongside the total flow time. The \ours{}
stage takes 34--152\,s per case and accounts for a median of about 4\% of total
runtime, ranging from 1.2\% to 9.3\%; placement alone takes 1.7--41
times as long.

\begin{figure*}[t]
 \centering
 \subfloat[Pin-3D]{\includegraphics[width=.49\textwidth]{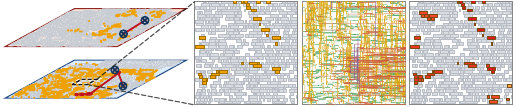}}\hfill
 \subfloat[\ours{}]{\includegraphics[width=.49\textwidth]{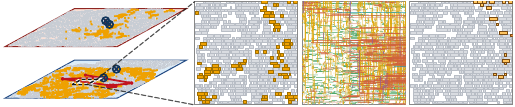}}
 \caption{JPEG/ASAP7 after routing, (a) Pin-3D and (b) \ours{} from the same
 placement state: the update reduces critical cells in the highlighted window.}
 \label{fig:layout}
\end{figure*}

\setcounter{topnumber}{2}\renewcommand{\topfraction}{.95}\renewcommand{\textfraction}{.05}
\suppressfloats[t]
\begin{figure}[t]
 \centering
 \sbox{\timinghistogrambox}{\hbox to\columnwidth{%
  \subfloat[Pin-3D reg2reg slack (ps)]{\includegraphics[width=.48\columnwidth]{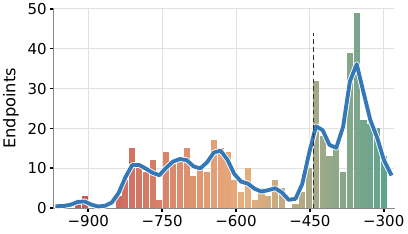}}\hfill
  \subfloat[ReSynP3D reg2reg slack (ps)]{\includegraphics[width=.48\columnwidth]{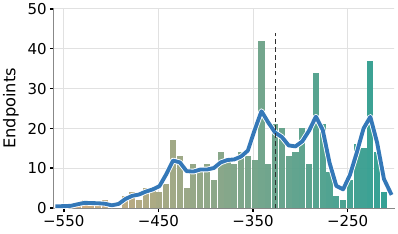}}}}
 \sbox{\timingablationbox}{\includegraphics[width=\columnwidth]{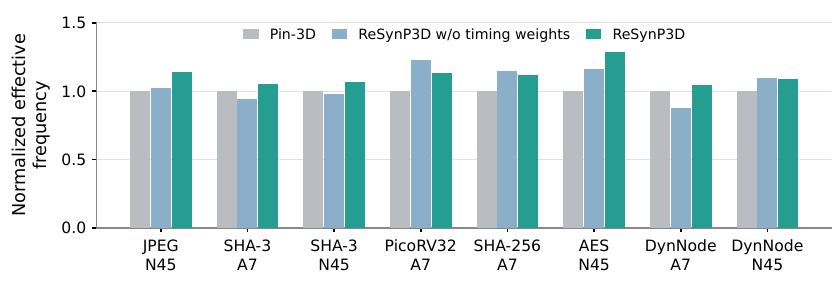}}
 \global\timingfigureheight=\ht\timingablationbox\relax
 \raisebox{\depth}{\resizebox*{!}{\timingfigureheight}{\usebox{\timinghistogrambox}}}\par
 \caption{Slack of the 500 worst paths, JPEG/ASAP7: shifted right.}
 \label{fig:slack-hist}
\end{figure}
\begin{table}[t]
\caption{Sensitivity to $K$ and $B$: post-route WNS (ps).}
\label{tab:kb}
\centering
\experimenttablefont
\begin{tabular*}{\columnwidth}{@{\extracolsep{\fill}}l*{10}{r}@{}}
\toprule
 & \multicolumn{5}{c}{ASAP7} & \multicolumn{5}{c}{Nangate45} \\
\cmidrule(lr){2-6}\cmidrule(lr){7-11}
 & \multicolumn{3}{c}{$K$ ($B=5\%$)} & \multicolumn{2}{c}{$B$ ($K=500$)} & \multicolumn{3}{c}{$K$ ($B=5\%$)} & \multicolumn{2}{c}{$B$ ($K=500$)} \\
\cmidrule(lr){2-4}\cmidrule(lr){5-6}\cmidrule(lr){7-9}\cmidrule(lr){10-11}
Design & 200 & 500 & 1000 & 2.5\% & 10\% & 200 & 500 & 1000 & 2.5\% & 10\% \\
\midrule
 JPEG & -640 & \textbf{-550} & -792 & -565 & -551 & -1837 & \textbf{-1505} & -2616 & -2128 & -2743 \\
 SHA-3 & -655 & \textbf{-557} & -638 & -774 & -557 & -509 & -443 & -437 & -503 & -562 \\
 PicoRV32 & 36 & -26 & 49 & 17 & 84 & -234 & -278 & -256 & -240 & -318 \\
 SweRV & 29 & \textbf{65} & 46 & 65 & 65 & -804 & -734 & -718 & -734 & -734 \\
 SHA-256 & -354 & -493 & -545 & -518 & -370 & -421 & -463 & -463 & -401 & -410 \\
 SystemC-AES & -10 & \textbf{-3} & -8 & -29 & -46 & -188 & -90 & -137 & -151 & -80 \\
 SHA-core & -80 & -78 & -91 & -89 & -72 & -183 & \textbf{-160} & -160 & -199 & -162 \\
 DynamicNode & -166 & -178 & -123 & -205 & -255 & -190 & -153 & -143 & -199 & -703 \\
\bottomrule
\end{tabular*}
\end{table}

\subsection{\ours{} Optimization Analysis}
\begin{figure}[t]
 \centering
 \includegraphics[height=\timingfigureheight]{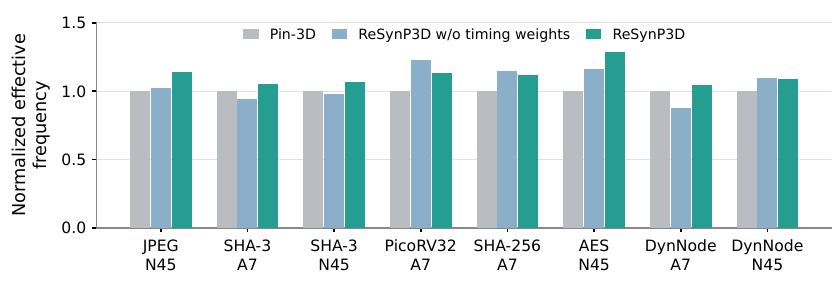}\par
 \caption{Timing-weight ablation: frequency normalized to Pin-3D.}
 \label{fig:lambda}
\end{figure}
\begin{table}[t]
\caption{Ablation study of partitioning methods.}
\label{tab:ablation}
\centering
\experimenttablefont
\begin{tabular*}{\columnwidth}{@{\extracolsep{\fill}}l*{12}{r}@{}}
\toprule
 & \multicolumn{6}{c}{ASAP7} & \multicolumn{6}{c}{Nangate45} \\
\cmidrule(lr){2-7}\cmidrule(lr){8-13}
 & \multicolumn{3}{c}{WNS (ps)} & \multicolumn{3}{c}{HBT} & \multicolumn{3}{c}{WNS (ps)} & \multicolumn{3}{c}{HBT} \\
\cmidrule(lr){2-4}\cmidrule(lr){5-7}\cmidrule(lr){8-10}\cmidrule(lr){11-13}
Design & ours & ILP & FM & ours & ILP & FM & ours & ILP & FM & ours & ILP & FM \\
\midrule
 JPEG & -550 & -348 & -392 & 1498 & 3431 & 3247 & -1505 & -3624 & -1183 & 1634 & 4013 & 4599 \\
 SHA-3 & -557 & -550 & -488 & 4769 & 6200 & 6708 & \textbf{-443} & -683 & -687 & 4655 & 5567 & 6628 \\
 PicoRV32 & -26 & -14 & 35 & 836 & 1290 & 1201 & \textbf{-278} & -391 & -366 & 1255 & 1934 & 1990 \\
 SweRV & \textbf{65} & 13 & -48 & 1597 & 2562 & 3437 & \textbf{-734} & -848 & -813 & 2186 & 3161 & 3884 \\
 SHA-256 & \textbf{-493} & -520 & -544 & 1426 & 1981 & 1837 & -463 & -477 & -463 & 1382 & 1873 & 1598 \\
 SystemC-AES & \textbf{-3} & -9 & -3 & 805 & 958 & 805 & \textbf{-90} & -190 & -133 & 790 & 916 & 793 \\
 SHA-core & \textbf{-78} & -80 & -81 & 218 & 482 & 605 & \textbf{-160} & -198 & -247 & 333 & 594 & 763 \\
 DynamicNode & -178 & -127 & -149 & 1029 & 1297 & 1541 & \textbf{-153} & -184 & -279 & 1215 & 1636 & 2171 \\
\bottomrule
\end{tabular*}
\end{table}

To examine whether placement-guided selection captures logic that remains
critical after routing, \cref{tab:overlap} compares the selected paths and
regions with post-route timing. Here $K$ reports the effective path count after budget trimming;
Kept and Top-100 count their endpoints that remain among Pin-3D's
post-route worst 500 and 100, respectively. Paths-base and Paths-ours
count the post-route worst 500 paths passing through the original region
in Pin-3D and the replacement region in \ours{}, respectively.
Original and Resyn give the corresponding cell counts before and after
resynthesis.
A median 86\% of the selected endpoints remain among Pin-3D's 500 worst
after routing. In 14 of 16 cases, 208--498 of these worst paths pass
through the selected region. These results indicate that placement-stage
timing identifies logic that remains relevant to post-route timing,
providing a useful target for resynthesis.
\Cref{fig:layout} contrasts the two flows on the routed layout, from left to
right: both dies, with the resynthesized region in orange, the worst path in
red with a marker at its tier crossing, and a window on the bottom die
that highlights a local concentration of critical logic;
the window as cells; the window as routed metal; and the window with each
cell shaded by the worst slack through it, on one scale from $-950$ to
$-211$\,ps. In the first three views the two arms are hard to tell apart:
the window is placed and routed alike under both flows. The last view is where
they differ. Under Pin-3D the window holds 45 cells of the 500 worst paths,
several at the red end of the scale; under \ours{} it holds 18 and none in
red, and on Pin-3D's worst endpoint the region's share of the critical delay
falls from 84\% to 26\%. This comparison suggests that restructuring
helps reduce the timing burden of logic in this window, consistent with
the opportunity illustrated in \cref{sec:prelim-resyn}.

\Cref{fig:slack-hist} shows that the gain is carried by the whole distribution:
each panel is the histogram of post-route slack over the 500 worst endpoints
of one flow, on one red-to-teal scale from $-970$ to $-200$\,ps, with the
median marked. Post-route WNS moves from $-950.1$ to $-550.2$\,ps. Pin-3D's
endpoints spread over 652\,ps (median $-443$\,ps); \ours{}'s span
339\,ps (median $-326$\,ps): the histogram shifts to the right as a body,
with a shorter negative tail, consistent with the post-route objective
of \cref{eq:problem} rather than only the local estimate
optimized in \cref{eq:map-target}.
Of the 239 endpoints both arms rank among their 500 worst, 233 improve.
Crossings per critical path remain similar, at 1.64 and 1.69, despite
the higher total HBT count in \cref{tab:main}. Thus, this case
illustrates how \ours{} can improve post-route timing without
requiring a reduction in the total number of tier crossings.

\subsection{Ablation and Sensitivity Analysis}
\label{sec:exp-ablation}

\Cref{tab:ablation} isolates the tier assignment of
\cref{sec:repartition}: using the same resynthesis recipes,
candidate-selection rule and physical flow, the replacement cells are
assigned by our local implementations of the PPA-aware ILP of~\cite{jeong2025ppailp}
and the bin-based FM tier partitioner of
Compact-2D~\cite{ku2018compact2d,panth2017shrunk2d}. Based on the displayed WNS values, the weighted
partition of \cref{eq:local-tier-cut} leads ILP in 12 of 16 cases
and bin FM in 9, with two additional ties against bin FM. The average
WNS advantages are about 162\,ps and 12\,ps.
It also achieves the lowest or tied-lowest HBT count in every case:
ILP and bin FM add about 767 and 1011 HBT connections per case
on average. ILP and bin FM also enforce area balance; \ours{}
instead weights nets by timing criticality.
\Cref{fig:lambda} removes the timing weights of \cref{eq:local-tier-cut}
($\lambda=0$, same partitioner): the assignment is unchanged in 8 cases; over the other 8 the
weighted cut gains 98\,ps on average and is ahead in 5, with HBT counts within 4 nets in 7 of them, so
the weights mainly change which nets are cut rather than how many.
For these eight cases, the plot compares Pin-3D, \ours{} without timing
weights, and the full method, with effective frequency
$F=1/(T_{\mathrm{clk}}-\mathrm{WNS}_{\mathrm{route}})$ normalized to Pin-3D.
The unweighted variant is ahead of Pin-3D in five cases, showing a
benefit without timing weights; the full method improves all eight.
This supports the criticality weighting in \cref{eq:net-criticality,eq:local-tier-cut}:
distinguishing timing-sensitive nets in the resynthesized candidate helps
retain gains through subsequent placement and routing that a uniform
crossing cost does not consistently preserve.

\Cref{tab:kb} assesses sensitivity to the default settings, $K=500$ and
$B=5\%$ of the design's cells. This configuration is the best or
within about 10\,ps of it in over half the cases. Each alternative loses more cases
than it wins, with median WNS changes from about
$-30$ to 0\,ps relative to the default. The sweep
also shows how the two parameters jointly control region size:
with $K=1000$ the budget clamps the selection in 14 cases, where the
region size stays within 8\% of the default, $B=10\%$ enlarges it by up to 2.2 times at an average WNS cost of
about 110\,ps per case, and $B=2.5\%$ halves it in 14 cases and loses 11.
Larger regions do not consistently improve timing, which supports the
budgeted selection in \cref{sec:extraction} and a shared default
across designs from 6.5k to 84k cells.

\section{Conclusion}
\label{sec:conclusion}
In this paper, we presented \ours{}, which couples placement-guided cross-tier logic
resynthesis with criticality-weighted local tier repartitioning between
consecutive placement iterations. Across eight designs
on two technology nodes, it improves post-route register-to-register WNS in all
16 reported cases, by 105\,ps on ASAP7 and 129\,ps on Nangate45 on average and by up
to 400\,ps. These improvements correspond to WNS-derived post-route
frequency gains of 9.1\% on average and up to 34.2\% over Pin-3D.

\section*{Acknowledgment}
This research was partially conducted by ACCESS-AI Chip Center for Emerging
Smart Systems, supported by the InnoHK initiative of the Innovation and
Technology Commission of the Hong Kong Special Administrative Region
Government. This work is jointly supported by the Research Grants Council
of Hong Kong SAR (Grant No.~T46-415/25-R).

\clearpage
\raggedbottom
\balance

\end{document}